# Microstructure evolution and thermal stability of equiatomic CoCrFeNi films on (0001) α-Al$_2$O$_3$

Younès ADDAB[1], Maya K. KINI[2], Blandine COURTOIS[1], Alan SAVAN[3], Alfred LUDWIG[3], Nathalie BOZZOLO[4], Christina SCHEU[2], Gerhard DEHM[2], Dominique CHATAIN[1*]

[1] Aix-Marseille Univ, CNRS, CINAM, Marseille, France
[2] Max-Planck-Institut für Eisenforschung GmbH, D-40237 Düsseldorf, Germany
[3] Chair for Materials Discovery and Interfaces, Institute for Materials, Faculty for Mechanical Engineering, Ruhr University Bochum, D-44801 Bochum, Germany
[4] MINES ParisTech, PSL - Research University, CEMEF – Centre de Mise en Forme des Matériaux, CNRS UMR 7635, 06904 Sophia Antipolis Cedex, France

[*] Corresponding author

**Abstract**

Homogeneous face-centered cubic (fcc) polycrystalline CoCrFeNi films were deposited at room temperature on (0001) α-Al$_2$O$_3$ (c-sapphire). Phase and morphological stability of 200 to 670 nm thick films were investigated between 973 K and 1423 K.
The fcc-phase persists while the original <111> texture of 30-100 nm wide columnar grains evolves into ~10 or ~1000 µm wide grains. Only the grains having specific orientation relationships to the sapphire grow; as in the case of pure fcc metal (M) films 4 orientation relationships (OR) are found: OR1 (M(111)[1$\bar{1}$0]//α-Al$_2$O$_3$(0001)[1$\bar{1}$00]) and OR2 (M(111)[1$\bar{1}$0]//α-Al$_2$O$_3$(0001)[11$\bar{2}$0]) and their twin-related variants (OR1t and OR2t). Below 1000 K, the film microstructure stabilizes into 10 µm wide OR1 and OR1t twin grains independent of film thickness. Above 1000 K, the OR2 and OR2t grains expand to sizes exceeding more than a 1000 times the film thickness.
Upon annealing, the films either retain their integrity or break-up depending on the competing kinetics of grain growth and grain boundary grooving. Triple junctions of the grain boundaries, the major actors in film stability, were tracked. Thinner films and higher temperatures favor film break-up by dewetting from the holes grooved at the triple junctions down to the substrate. The grain boundaries of the OR2 and OR2t grains migrate fast enough to overcome the nucleation of holes from which break-up could initiate. The growth of the OR2 and OR2t grains in this complex alloy is faster than in pure fcc metals at equivalent homologous annealing temperatures.

# 1. Introduction

First studies of multi-principal-element alloys (MPEAs), earlier referred to as "high entropy alloys" (HEAs), were published in 2004 [1-4]. These alloys, at least quaternary ones, are complex solid solutions in which the different components occupy the available lattice sites with the same probability [5,6]. This new class of materials is widely studied owing to their promising physical properties with potential applications [7-10], and the need to understand their formation, their stability and the origin of their unusual properties. The stability of the single-phase multicomponent alloys was originally supposed to be related to a large configurational entropy [11]. Nowadays, intrinsic properties other than entropy, are considered to explain such stability [12-15]. This is supported by the recent measurement of the excess entropy of the CoCrFeNi alloy which is zero above 300 K while having a high maximum at 100 K [16].

This paper focuses on the equiatomic CoCrFeNi alloy. Above 873 K, it has a face-centered cubic (fcc) structure as shown both experimentally [6,17-21] and by numerical and first principle calculations of phase diagrams [21-25]. The fcc CoCrFeNi remains stable after annealing treatments up to 1373 K, as shown in [18] (1 h at 1273 K) and in [20] (1 h at 1373 K). The calculated melting temperature ($T_m$) of CoCrFeNi is 1707 K [26], consistent with the measured value of 1717 K [27]. Below 630 K, the σ-phase is stable [21] but there is a lack of reliable data on phase equilibrium below 600 K. Calculated phase diagrams report that when Cr exceeds 25%, a Cr-rich σ-phase precipitates [15,21].

The present work is dedicated to thin films of this specific quaternary MPEA deposited on single-crystalline substrates. Thin films provide a unique opportunity to control the homogeneity of complex alloys, which in turn facilitates the study of their phase stability. On the other hand, the use of single-crystal oxide substrate enables the microstructure of the film to be controlled. It is therefore easier to understand some of the properties of this new category of alloys in thin films than in bulk materials which have a complex microstructure and are more likely to lack homogeneity. Also, thin films facilitate the studies of MPEA/HEA thermodynamic stability since their defects like surfaces, interfaces and grain boundaries, which enhance the kinetics and nucleation of phase transformation as shown in [28] for the bulk Cantor alloy (CrCoFeMnNi), are easier to track.

Different techniques can be used to prepare metallic alloy thin films, such as sputter deposition [29,30], pulsed laser deposition [31], thermal spraying [32], electrodeposition [33], among others. The reviews published on MPEA/HEA films contain discussions of preparation methods, composition design, properties and potential applications [34,35]. In the present work, thin films were prepared by magnetron sputtering, a technique which has been successfully used for rapid atomic-scale investigations of the phase evolution of MPEA/HEA films, i.e., stability against decomposition into competing phases [36] and oxidation [37].

The morphological integrity of annealed thin solid films is a complex issue. For a given film composition, morphology may change depending on several parameters such as substrate preparation, annealing time, heating temperature, porosity and film thickness. Several studies on polymer [38-40] and pure and alloyed metallic [41-43] polycrystalline films have shown that the thinner the film, the lower the temperature at which break-up takes place. Polycrystalline metallic films on oxides are usually unstable under annealing, and break-up [44] well-below the melting point of the film. The main stages of break-up are hole nucleation, hole growth, hole percolation and dewetting into islands. Holes form either by grooving at grain boundaries and their triple junctions from the free surface towards the interface with the substrate, or by nucleation of voids at the interface which grow towards the free surface [44-47]. Grooving is the most critical process in film break up. However, grain growth stabilizes the film morphology by eliminating grain boundaries and triple junctions. Thus, the film integrity depends on a complex interplay between the kinetics of grain boundary and triple junction grooving and the kinetics of grain growth.

This paper presents an original study of the evolution of <111>-textured, quaternary, equimolar fcc CoCrFeNi films deposited on c-sapphire ((0001) α-$Al_2O_3$ single-crystal), annealed up to 1423 K (0.8 $T_m$, where $T_m$ stands for the melting temperature). It provides information on the thermodynamic stability of the fcc phase, on the orientation relationships between the grains of the film and the substrate and on the morphological changes of these films upon annealing. The microstructure evolution is analyzed by tracking the film triple junctions at which holes nucleate; it allows a better understanding of the complex interplay between film texture, film thickness and annealing temperature. Unexpectedly, these alloy films behave like pure fcc metal films on c-sapphire. The complex chemistry of the film has virtually no impact on the evolution of its microstructure, other than to promote its morphological stability through the rapid growth of very large grains of a particular orientation relationship. The interesting feedback from this study is that the conclusions on the behavior of these fcc alloy films can be extended to thin films of pure fcc metal on c-sapphire.

## 2. Experimental

Equimolar polycrystalline CoCrFeNi films were deposited on ∼ 1 cm x 1 cm pieces of sapphire, cleaved from wafers (5.08 cm in diameter, 400 μm thick, provided by CrysTec GmbH), with a c-plane surface with less than 0.1° of miscut. Substrates were prepared in a clean room to prevent contamination by air-suspended dust particles. Since the sapphire dust produced by the cleavage can accelerate dewetting during sample annealing, each substrate piece was further cleaned in two ultrasonic baths (5 min in acetone, 5 min in isopropanol), then rinsed in de-ionized water and dried under nitrogen flow. Finally, the substrates were exposed to an oxygen plasma at 473 K for 10 min in a barrel reactor (Nanoplas, France) to burn off organic contaminants.

Films were co-deposited from four elemental sources in a specialized magnetron sputter deposition system (DCA, Finland) [48] at room temperature. Today, it is the most advanced tool to fabricate homogeneous multi-component films. Individual 100 mm diameter targets of Cr (99.95 wt.%, MaTecK, Germany), Fe (99.99 wt.%, Evochem, Germany), Co (99.99 wt.%, MaTecK) and Ni (99.995 wt.%, K.J. Lesker, USA) were confocally arranged. The sapphire substrates were transferred into the deposition chamber with a base vacuum of $4.5 \times 10^{-6}$ Pa via a load lock. Prior to deposition, all targets were pre-cleaned by sputtering against closed shutters, while the substrates received a low power radiofrequency (8-10 W RF (13.56 MHz)) Ar ion bombardment for 6 min to remove adsorbed species. The power applied to each cathode was adjusted to achieve the desired elemental composition of the co-sputter-deposited thin films, with an overall growth rate of 0.13 nm/sec. During deposition, the substrate table was rotating at 20 revolutions per minute to achieve a uniform composition, later confirmed by energy dispersive X-ray spectroscopy (EDS) in a scanning electron microscope (SEM). A gradual temperature rise to 311 K at the end of the deposition process was indicated by a thermocouple inside the substrate manipulator. Measured film thicknesses were 200 nm, 270 nm, 500 nm and 670 nm. Their compositions as determined by SEM-EDS are summarized in Table 1. All films are equimolar within an accuracy of 1 at. %. Three EDS measurements of the same 500 nm thick film, done on two different instruments, showed consistent values within ±0.7 at. % (standard deviation), that indicates a reliable composition.

Finally, the samples were transferred through air into an alumina tube fitted in a tubular furnace, and heat treated under a 10 $cm^3$/min gas flow of 60% Ar + 40% $H_2$. Anneals were performed at 973 K, 1273 K or 1423 K for 1 h, each one on a different as-deposited samples. The heating rate was 25 K/min while the cooling rate was imposed by switching the furnace power off as described in detail in [49]. It takes about two hours for the temperature decreases down to 723 K where the diffusion of the alloy components becomes very slow.

Table 1: Compositions of the as-deposited alloys measured by EDS with the standard deviation of repeated measurements.

| film thickness | Co (at. %) | Cr (at. %) | Fe (at. %) | Ni (at. %) |
|---|---|---|---|---|
| 200 nm | 24.2±0.4 | 25.1±0.1 | 26.3±0.3 | 24.4±0.2 |
| 270 nm | 22.3±0.1 | 23.5±0.1 | 27.1±0.1 | 27.1±0.1 |
| 500 nm | 24.2±0.6 | 26.0±0.6 | 26.7±0.7 | 23.0±0.6 |
| 670 nm | 22.0±0.2 | 23.6±0.3 | 27.7±0.2 | 26.6±0.3 |

As-deposited and annealed films were characterized using SEM in the secondary electron (SE) imaging mode, EDS, electron backscattered diffraction (EBSD) and X-ray diffraction (XRD). SEM images and EBSD data were acquired at 10-20 kV on a Zeiss SUPRA40 field emission gun SEM equipped with a Bruker EBSD detector. EBSD data were processed using the Esprit 2.1 software package from Bruker and the MTeX Matlab toolbox [50]. XRD measurements were performed on two different systems: a Panalytical X'pert Multipurpose Diffractometer operating with Cu-K$_{\alpha1}$ radiation (0.154 nm), and a step size of 0.03 degrees, and a Seifert X6_WS diffractometer operating with Co-K$_{\alpha1}$ radiation (0.178 nm), and a step size of 0.03 degrees.

Since triple junctions are the major source of film instability [51,52], the microstructural evolution of the polycrystalline CoCrFeNi film has been tracked by measuring the density of triple junctions (depressions or holes) revealed by a darker contrast on the SE images.

It is worth noting that the nominally 40% H$_2$ present in the annealing atmosphere did not prevent Cr oxidation but significantly slowed down chromium oxide growth at the surface of the samples. An oxide layer has been observed on the samples after annealing, due to the presence of oxygen traces in the gas flow. The grown oxide, a few tens of nanometers thick, is thin enough that the surface maintains a metallic shine. However, it had to be removed by Ar ion-polishing before EBSD data acquisition on the alloy films.

## 3. Results
### 3.1. Fcc phase stability and <111>-texture

The pristine surfaces of the films of different thicknesses (200 nm, 270 nm, 500 nm and 670 nm) are similar. Figure 1 displays the as-deposited film morphology of a 500 nm thick CoCrFeNi film. The top-view in Fig. 1a shows that the film is comprised of grains with lateral sizes between 30 and 100 nm; similar microstructures have been reported for single-element and binary alloy (Au, AuPt, Ni, Al, Cu) [42,49,53-55] or MPEA (AlCoCrCuFeNi) [56] films deposited at room temperature. At lower magnification, the surface of the film appears continuous and uniform. The lighter vertical band displayed on the right of Fig. 1a is a frequent topographic anomaly of our films where the grains are slightly above the nominal surface level. Fig. 1b is a SE image of the cross-section produced by focused ion beam (FIB) milling. The film appears compact and homogeneous through the thickness. The vertical features indicate a columnar grain structure. The interface between the film and the substrate appears to be sharp at the scale of the observation.

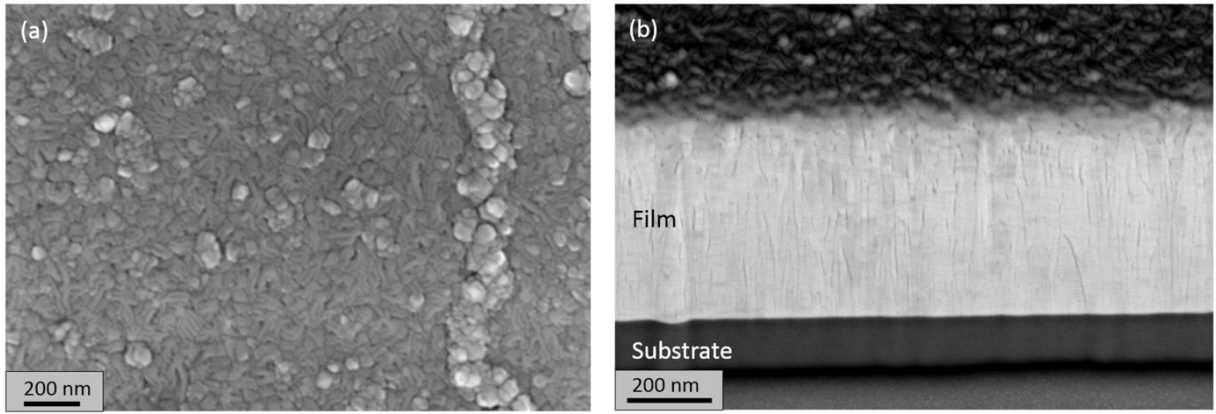

Figure 1: SE images of a 500 nm thick as-deposited CoCrFeNi film: (a) top view of the nanometric grains (b) cross section (inclined view) showing the columnar grain structure.

The stability of the fcc CoCrFeNi phase has been analyzed by SEM-EDS and XRD, before and after annealing. Polycrystalline thin films are interesting to investigate phase diagrams because they contain grain boundaries and surfaces on which new phases can nucleate after a short diffusion path upon annealing.

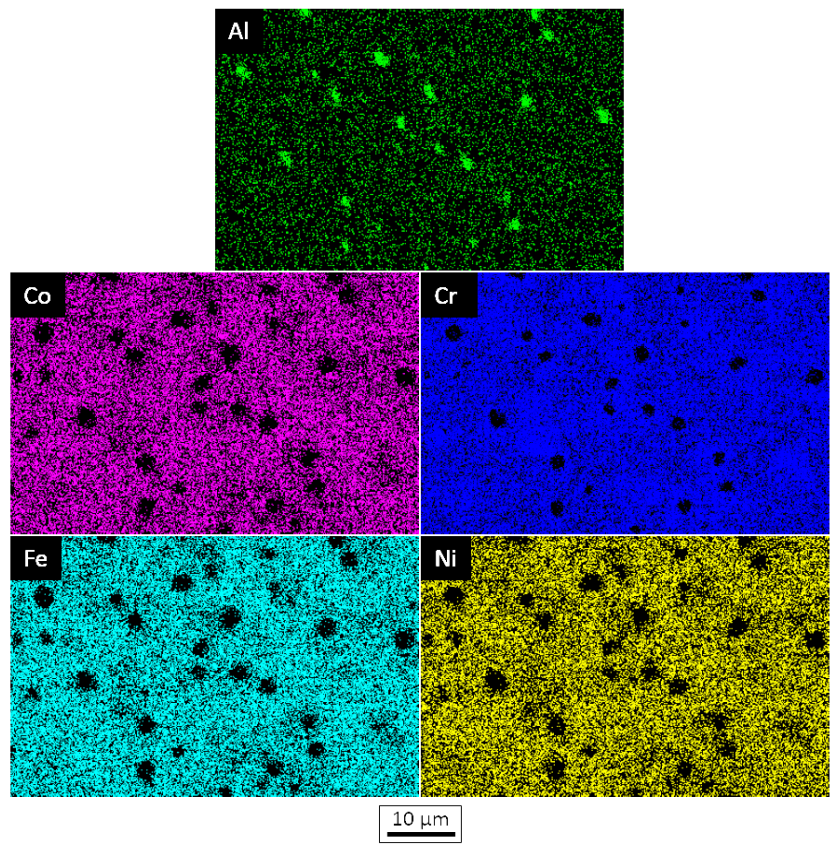

Figure 2: SEM-EDS maps of a 670 nm CoCrFeNi film deposited on c-sapphire, annealed for 1 h at 1423 K under an Ar-40% $H_2$ atmosphere. The bright regions on the Al map reveal the substrate at holes formed during annealing and coincide with the black regions on the Co, Cr, Fe and Ni maps.

Figure 2 presents the EDS maps of Al (from sapphire substrate) and the four elements of the CoCrFeNi alloy, for a 670 nm film annealed for 1 h at 1423 K. The regions where the Al signal is high (brighter regions) and where neither Co, Cr, Fe nor Ni is detected correspond to the sapphire substrate at the bottom of the holes running through the film (see details in § 3.3 and Fig. 6i). These elemental distribution maps show that at the length scale of SEM-EDS there is no phase separation occurring in the film or associated with the hole formation. Areas of higher Cr content (see the brighter areas on the Cr map of Fig. 2) are likely to be due to the local formation of chromium oxide at the surface of the film; this does not alter the concentration distribution of the three other components as already mentioned in the paper by He et al. [17] which shows a uniform elemental distribution in SEM-EDS maps of bulk CoCrFeNi annealed at 1023 K for 800 h in air.

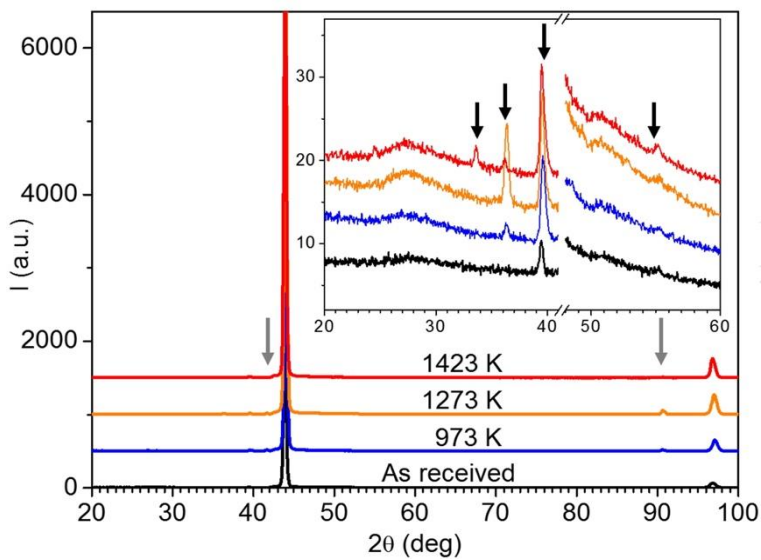

Figure 3: XRD diagrams measured on as grown 670 nm CoCrFeNi films and on annealed ones, for 1 h at 973 K, 1273 K and 1423 K under an Ar - 40% $H_2$ atmosphere. The main frame displays the location of two weak peaks due to sapphire (see grey arrows) and two more intense ones from fcc CoCrFeNi (see details in the text). Arrows in the inset indicate the position of expected $Cr_2O_3$ peaks.

Figure 3 shows XRD patterns measured on as-deposited 670 nm films and on films annealed for 1 h at 973 K, 1273 K and 1423 K. Peaks from the c-sapphire substrate appear at 41.7° and 90.5° for the (0 0 0 6) and (0 0 0 12) planes, respectively; they are very weak and the former is barely visible. The two main peaks in all XRD patterns, before and after annealing, appear at 43.9° and 96.7°. They correspond to the {111} and {222} planes of the fcc CoCrFeNi and are the only peaks of this phase appearing on the XRD patterns. This indicates that: (1) the as-deposited films are fcc and textured, with the {111} planes parallel to the (0001) plane of the sapphire substrate, and (2) upon annealing the fcc phase is stable and the microstructure remains <111>-textured. Other very weak peaks (see inset on the right of Fig. 3) are present in the diffractograms, associated with $Cr_2O_3$ formed at the surface of the film. Such oxidation slightly depletes the alloy from Cr but this does not impact the stability of the fcc phase as reported in [17,21,24].

The lattice parameter of the deposited CoCrFeNi extracted from the XRD data is 0.357 nm, in agreement with the values given in literature: 0.356 nm [6] and 0.359 nm [57].

**3.2. Grain size and grain orientation relationships on c-sapphire**

This section shows that grain growth in the annealed films occurs with different kinetics depending on the orientation relationship of each grain to the c-sapphire substrate (see Table 2).

The size of the grains and their orientation relationships (ORs) within the <111>-textured fcc film on c-sapphire substrate were analyzed using XRD and SEM imaging for the as-deposited film which has nano-grains, and using EBSD for the annealed samples with grains larger than a micrometer. The ORs of the grains of the metallic film with the c-sapphire are obtained by comparing the {110} pole figures of the fcc-alloy to the stereogram of the sapphire substrate, centered on the (0001) pole, and displaying the {1$\bar{1}$00} {11$\bar{2}$0} and {1$\bar{1}$02} poles. Both crystallographic data on the film and the sapphire are acquired during the same run on each film-sapphire sample as explained in [49].

As previously shown by XRD, all films are <111>-textured, i.e. all their grains have a <111> direction perpendicular to the substrate surface. Their preferred in-plane orientation to the c-sapphire depends on whether the film is as-deposited or annealed. The grains of the annealed films adopt the same two ORs as the ones reported for pure fcc metals (M) on c-sapphire [55,58-62]. They are called OR1 and OR2, have two twin-related variants each and may be written as:

OR1                      M(111)±[1$\bar{1}$0]//α-Al$_2$O$_3$(0001)[1$\bar{1}$00]

OR2                      M(111)±[1$\bar{1}$0]//α-Al$_2$O$_3$(0001)[11$\bar{2}$0]

The ± indicates the two twin-related variants OR1 and OR1t, and OR2 and OR2t, respectively. OR1t (OR2t) is rotated from OR1 (OR2) by 60° around the <111> normal to the film surface. OR1(OR1t) and OR2(OR2t) are rotated by 30°. The notation OR1/t (OR2/t) will be used hereafter and in Table 2 to refer to the coexistence of OR1 and OR1t (OR2 and OR2t). Table 2 gathers the texture, microstructure and morphology details of all films. The microstructure of the annealed films will be discussed in detail in a later section §3.3.

Table 2: Summary of size (estimated from SE or EBSD images) and orientation relationship of the grains in the 200, 270, 500 and 670 nm thick CoCrFeNi films on c-sapphire, as-deposited and annealed at 973 K, 1273 K and 1423 K for 1 h.

|  | Film thickness (nm) | Grain size (μm) | Preferred ORs | Morphology |
|---|---|---|---|---|
| as-deposited | 200-670 | 0.03 - 0.1 | OR1/t | smooth surface |
| 973 K | 200-670 | ≤1 | OR1/t | film with depressions |
| 1273 K | 670 | ~10 | OR1/t | film with depressions |
|  | 500 | ~1000 | OR2/t | film with depressions |
|  | 270 | ~10 | OR1/t | network of connected crystals |
|  | 200 | ~1000 | OR2/t | film with holes in grains |
| 1423 K | 670 | ~10 | OR1/t | film with holes |
|  | 500 | ~1000 | OR2/t | film with holes in grains |
|  | 270 | ~10 | OR1/t | poly-crystalline fingers |
|  | 200 | 1 - 3 | OR1/t | single-crystalline islands |

All the as-deposited films of different thicknesses and the ones annealed at 973 K have small grains which lateral size does not exceed 1 μm. Figs. 4a and 4c display the XRD {110} pole figure of an as-deposited 670 nm film and the sapphire stereogram of the substrate, respectively. Fig. 4a shows that the azimuthal orientation of the grains in the as-deposited <111>-textured film is not random; the grains prefer to be in OR1 and its twin variant OR1t (shown in Figs. 4d and 4e). OR1/t is also preferred in all the other as-deposited films, with a similar frequency of the twins OR1 or OR1t. The as-deposited films have about 4 vol.% of OR2/t grains whatever the thickness of the films, as measured on XRD pole figures. This is an important feature since the growth of grains with OR2/t is very likely to proceed by overgrowth of already existing OR2/t grains at the expense of grains of other orientations. This can be inferred from a recent study of a <111>-textured Al film on c-sapphire with OR1/t and OR2/t grains, which upon annealing close to the melting point of Al, showed that OR2/t grains grow at the expense of OR1/t grains [58].

Since after annealing at 973 K, the grains of all films have grown to sizes in the micrometer range, their orientations on c-sapphire can be analyzed using EBSD. This annealing favors the growth of the OR1/t grains which are now the only ones to be present on the EBSD {110} pole figure of Fig. 4b. Note that XRD (Fig. 4a) reports highest intensities for OR1 and OR1t, but still some intensities for OR2 and OR2t as well. The preference for OR1/t has already been observed for pure fcc metals like Al [58,59], Cu [55,60,61] and Pt [62] grains on c-sapphire. The absolute values on the intensity scale in the pole figures presented on Figs. 4a and 4b cannot be directly compared since the pole figures were built from crystallographic data acquired using different techniques: XRD and EBSD, respectively. In both Figs. 4a and 4b it is observed that one of the OR1 is preferred to its twin.

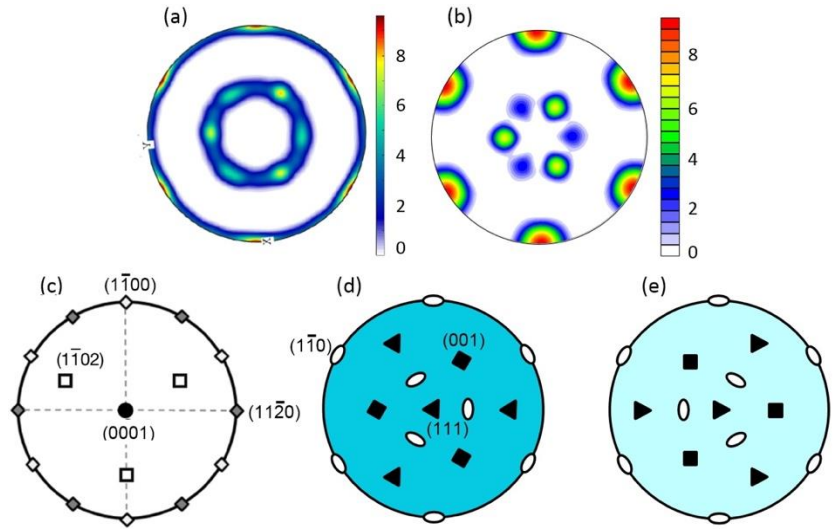

Figure 4: (a) XRD {110} pole figure of a 670 nm thick CoCrFeNi film as-deposited on c-sapphire; (b) EBSD {110} pole figure of the grains of a 670 nm thick metallic film on c-sapphire annealed for 1 h at 973 K showing the distribution of the {110} poles. The intensities are given in multiples of a random distribution (MRD). (Area scanned: 40 μm x 100 μm); (c) sapphire stereogram centered on the (0001) pole and containing the poles of the $\{1\bar{1}00\}$, $\{11\bar{2}0\}$ and $\{1\bar{1}02\}$ planes shown as white and dark lozenges, and squares, respectively. (d) and (e) are the stereograms of fcc metal on sapphire in OR1 and OR1t, respectively. OR1 and OR1t are identical ORs as observed from the symmetrical positions of the poles in the sapphire and the fcc metal stereograms.

In the thicker films annealed at 1273 K and 1423 K grains have grown and have adopted either OR1/t or OR2/t as indicated in Table 2. The EBSD maps of the 670 nm thick film show that the grains have grown to about 10 μm and have adopted mostly OR1/t, while the grains in the 500 nm thick film have grown "abnormally" to sizes 1000 times larger than the film thickness and have adopted OR2/t.

Figure 5 illustrates the correlation between the grains in OR2/t on c-sapphire and their sizes by using the EBSD data obtained on the 500 nm thick film annealed at 1273 K. Fig. 5a is an EBSD map acquired on this film and Fig. 5b is the corresponding {110} pole figure. The color codes of the map and the pole figure are identical. The film has been scratched on purpose (black regions of the map of Fig. 5a) to acquire EBSD data on the c-sapphire and the film during the same run. Comparison of the {110} pole figure of the alloy with the sapphire stereogram of Fig. 5c shows that the in-plane orientations of the grains of the film correspond to OR2 or OR2t within ±10° each. The stereograms of these two fcc crystal orientations are displayed in Figs. 5d and 5e, aligned with the substrate stereogram reminded on Fig 5c. Fig. 5a shows that the very large grains extend over several hundreds of μm and up to a mm, and that the two grain variants cover equivalent proportions of the scanned area: 48% for light purple grains and 52% for dark purple grains. This feature has been checked on other regions of the sample.

Such giant grains are remarkable considering that literature reports that in thin films grain growth stagnates at sizes of about 20 times the film thickness [53,63–65]. The OR2/t-type grains have grown at the expense of others which had unfavorable ORs with the substrate as was observed for Al films on c-sapphire [58]. While less frequently, OR2/t has also been previously observed in other pure fcc metals (Cu, Pt, Ir) on c-sapphire [60,62,66].

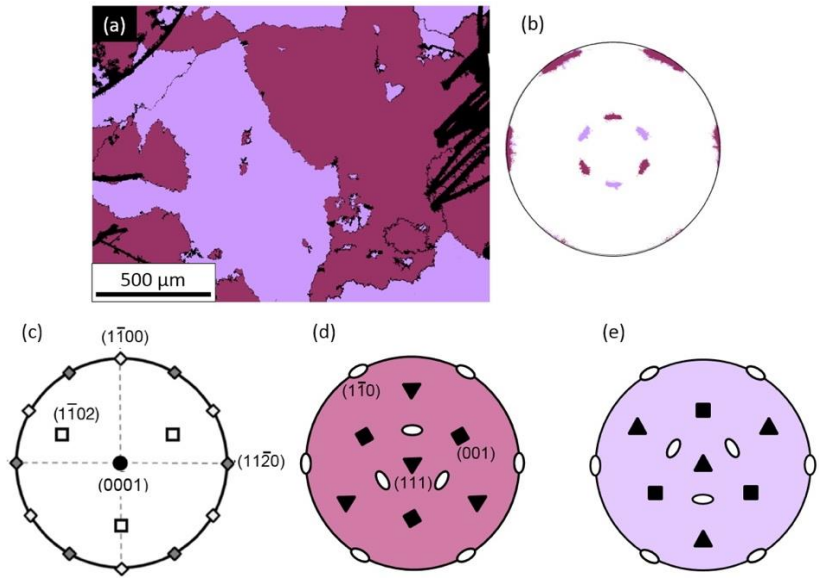

Figure 5: (a) EBSD map and (b) the corresponding {110} fcc pole figure of a 500 nm CoCrFeNi film annealed for 1 h at 1273 K, the grains of the map and their {110} poles on the pole figures have same colors; (c) the stereogram of sapphire (same as the one used in Fig. 4c) built from the EBSD data acquired in the black region of (a) where the film has been scratched. (d) and (e) are the stereograms of fcc grains positioned in OR2 and OR2t on the sapphire, respectively. Their background is colored according to the two OR2 variants. OR2 and OR2t are different OR as seen when comparing the positions of the poles inside the stereograms of sapphire with those inside the stereograms of the fcc alloy. Note in (b) the spread in OR2 and OR2t, which is different from the case of OR1 and OR1t (Fig. 4b).

Upon annealing at 1273 K and 1423 K, the 270 nm film behaves like the 670 nm film as far as the size and OR of the grains are concerned. However, at these temperatures, the 270 nm thick film breaks up because it is too thin; at 1273 K, (see Fig. 6e), holes in the film expand along the network of the grain boundaries, and at 1423 K, the holes have percolated (see Fig. 6f) and left wavy pearled strings of OR1 and OR1t grains.

The 200 nm thinner film behaves differently. At 1273 K, the film has giant OR2/t grains with holes within the grains (see Fig. 6b), while at 1423 K the film has broken-up into single-crystals (see Fig. 6c). A large fraction of them are in OR1/t, but 22.4±9.6% are in OR2/t. This mean ratio and its standard deviation have been obtained by analyzing the OR of 562 crystals distributed on 20 SE images such as the one shown in Fig. 6c; the ORs have been determined by comparing the <110> directions of these facetted crystals in their {111} interfacial plane (i.e. the direction of the edges of the top {111} facet of each crystal like the one shown in the inset of Fig. 6c) to the <1$\bar{1}$00> in-plane directions of the sapphire measured by EBSD. For these annealing conditions, the OR2/t grains size has been estimated to be 10 µm, based on the spatial distribution of the islands while the OR1/t grains have smaller sizes in the 1-3 µm range (see Table 2).

This above series of data clearly shows that there is a correlation between the size of the grains and their ORs. As will be shown in detail in the next section, the morphology of the film does not simply depend on the film thickness, but also on the microstructure resulting from the competition between grain growth and grain boundary grooving, which in turn depends on the annealing temperature.

### 3.3. Evolution of the film morphology after a single anneal

Figure 6 presents the morphology of films of the four different thicknesses after a single 1 h annealing at three different temperatures: 973 K, 1273 K and 1423 K. At this stage of the presentation of the results, the 200 and 270 nm films will be referred to as the "thinner films" and the 500 and 670 nm films will be referred to as the "thicker films". Images (a) to (c) of Fig. 6 correspond to the 200 nm thickness films, images (d) to (f) correspond to 270 nm thickness films, images (g) to (i) correspond to 500 nm thickness films and images (j) to (l) correspond to 670 nm thickness films. It is important to emphasize that each annealing treatment has been performed using a different as-deposited sample; thus, the images of Fig. 6 show "single-anneal" morphologies for each sample thickness rather than an annealing sequence.

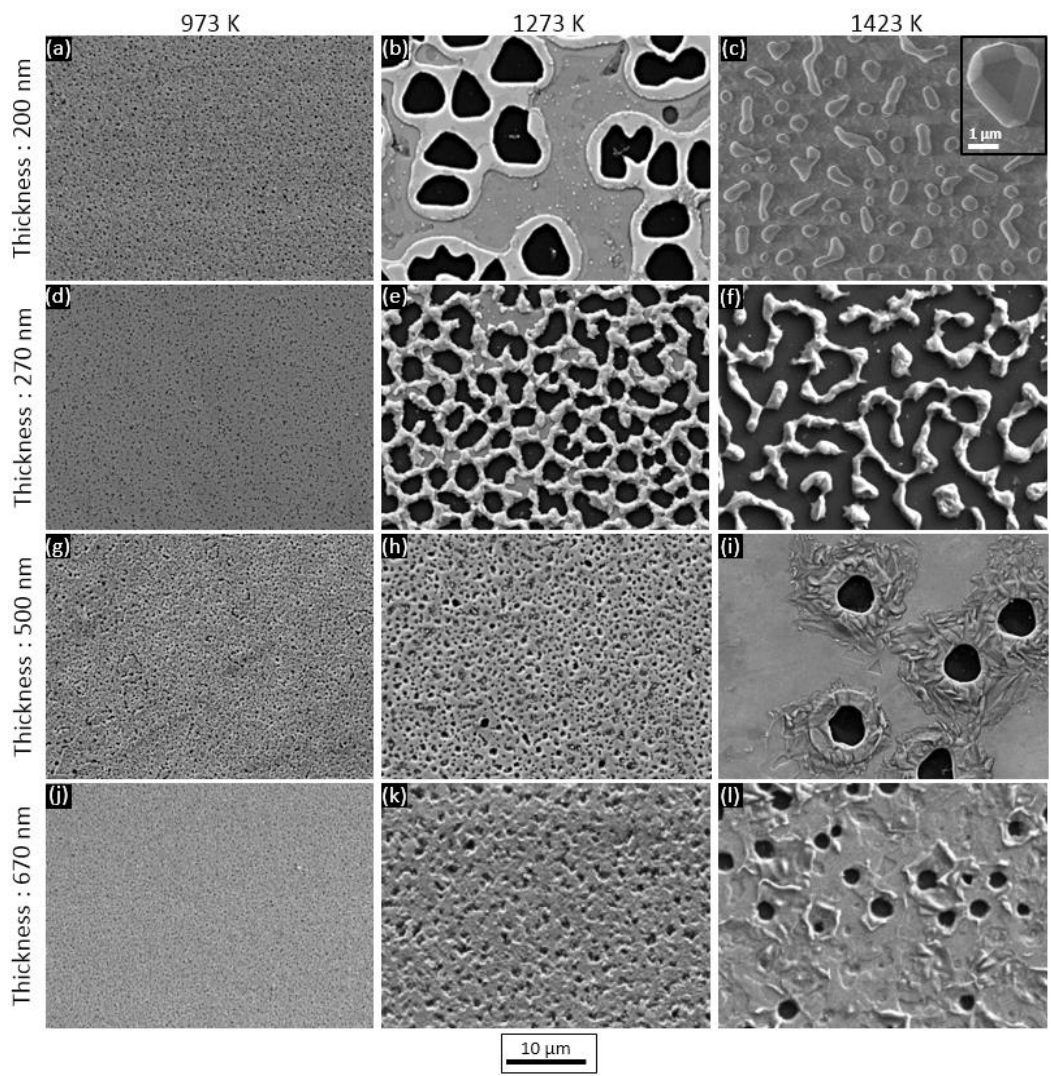

Figure 6: SE surface images of CoCrFeNi films annealed for 1 h under an Ar - 40% $H_2$ atmosphere, displayed at the same magnification. (a)-(c): 200 nm thick films annealed at (a) 973 K, (b) 1273 K and (c) 1423 K where the inset shows a facetted crystal; (d)-(f): 270 nm thick films annealed at (d) 973 K, (e) 1273 K and (f) 1423 K, (g)-(i): 500 nm thick films annealed at (g) 973 K, (h) 1273 K and (i) 1423 K, (j)-(l): 670 nm thick films annealed at (j) 973 K, (k) 1273 K and (l) 1423 K.

Upon annealing, grain growth and grooving at triple junctions and at grain boundaries undergo simultaneously [44]. When grains grow, their boundaries move to reduce their total area and consequently, triple junctions move and/or disappear. The groove pits which had begun to form disconnect from the underlying triple line/grain boundary connected to the substrate. These tips get blurred and turn into depressions which eventually heal by surface diffusion. Thus, in principle, grain growth prevent film break-up by eliminating triple junctions and grain boundaries. This ensures the morphological stability of the film. The conducted experiments allow to find the conditions for which grain growth overcomes triple junction grooving, which is the fastest diffusional process to produce holes in polycrystalline film [51].

Table 3: Pit size (with the standard deviation of measurements), density and area fraction for the 200, 270, 500 and 670 nm thick CoCrFeNi films annealed at 973 K, 1273 K and 1423 K. No measurement has been performed for the two thinnest films annealed at the highest temperature since they have broken-up into islands.

| Temperature (K) | Film thickness (nm) | Pit size (µm) | Pit density (/mm²) | Pit area fraction |
|---|---|---|---|---|
| 973 | 670 | 0.20±0.05 | 2.3E+06 | 0.07 |
| 973 | 500 | 0.16±0.09 | 7.7E+06 | 0.15 |
| 973 | 270 | 0.19±0.07 | 3.2E+06 | 0.09 |
| 973 | 200 | 0.20±0.05 | 7.1E+05 | 0.02 |
| 1273 | 670 | 0.98±0.30 | 2.0E+05 | 0.15 |
| 1273 | 500 | 0.56±0.20 | 9.9E+05 | 0.24 |
| 1273 | 270 | 2.96±0.86 | 5.9E+04 | 0.40 |
| 1273 | 200 | 5.32±1.30 | 1.5E+04 | 0.34 |
| 1423 | 670 | 2.00±0.50 | 1.5E+04 | 0.05 |
| 1423 | 500 | 4.94±0.41 | 1.5E+03 | 0.03 |

The two thinner films have broken-up at the highest annealing temperature (Figs. 6c and 6e). All the other annealed films remain continuous. The SE images of Fig. 6 show that the film contain scattered darker spots which correspond to pits either in the shape of depressions or holes running through the entire film thickness. We have extracted the projected areas and the sizes of these dark spots from the images of Figs. 6 and other similar ones. The values presented in Table 3 are averaged from at least three different images covering between 10 µm² and 9x10$^5$ µm², depending on the size of the dark spots (equivalent circle diameter between 0.16 and 5.32 µm). The contrast threshold for spot detection has been carefully chosen on each image after a manual check on representative small areas.

### *3.3.1. Hole nucleation: triple junction grooving by surface diffusion*

This section addresses pit nucleation at triple junction of the films which takes place during the so-called incubation time [44] before holes have formed through the entire thickness of the film. We focus on the films annealed at 973 K and of the two thicker films annealed at the intermediate temperature (1273 K).

After the 973 K anneal, the films display evenly distributed dark spots corresponding to surface depressions (Figs. 6a, 6d, 6g and 6j). They have similar sizes between 0.16 ± 0.09 µm (Fig. 6g) and 0.20 ± 0.05 µm (Fig. 6j), their density is on the order of 10$^6$ per mm² and their projected area occupies between 2% and 15% of the total film area. It is likely that those spots correspond to grooves formed at grain boundary triple junctions by surface diffusion at 973 K [51].

Similar depressions have formed in the two thicker films annealed at 1273 K (Figs. 6h and 6k). They have larger sizes of 0.56 ± 0.20 µm (Fig. 6g) and 0.98 ± 0.30 µm (Fig 6j), in the 500 and 670 nm thick film, respectively. They have smaller densities on the order of $10^5$ per mm², and their projected surfaces occupy 20 ± 5 % of the total film area. The depressions are about 10 times less frequent after annealing at 1273 K than at 973 K, suggesting that the grains of the thicker films have grown larger at the higher temperature. This is due partly to grain growth and partly to the existence of a critical size, below which the depressions heal and above which they expand into holes running all through the film thickness upon annealing [52,67]. Below the critical size, a grain boundary in motion can still detach from a grooved triple junction and prevents the depressions of growing further to form a hole running through the film thickness. Once there is no more triple line attached to the root, the groove heals.

### 3.3.2. Hole growth and dewetting mechanisms

This section focuses on the second step of the morphological change of polycrystalline films, *i.e.*, the deepening of the triple junction groove down to the substrate, and the growth of the holes thus formed through the entire thickness of the film. Here, the kinetics of dewetting is not under focus; instead we are interested in identifying the path through which film break-up happens by examining the shape of the dewetting holes.

The dewetting mechanism of a film starting from the holes running through it, depends on the location of these holes in the film which, in turn, depends on its microstructural evolution. When the grains grow rapidly to a large size, the grain boundaries detach from the hole formed through the film at former triple junctions. Such holes are then isolated within single grains and expand according to the dewetting mechanism observed in single crystals, with facetted shapes depending on the crystal symmetry (see e.g. Fig. 6b) [44,68,69]. On the other hand, grain boundaries can remain attached to the hole because grain boundary grooving has impeded their migration. Such holes expand into the film along the grain boundary network [46], forming channels which eventually percolate and transform the film into a connected network of crystals (see e.g. Fig. 6f). This dewetting mechanism is typical for polycrystalline films. The two thinner films annealed at 1273 K and the two thicker films annealed at 1423 K are shaped by one or both dewetting mechanisms.

Hole dewetting in single-crystal grains was observed in the 200 nm film at 1273 K (Fig. 6b) and the 500 nm film at 1423 K (Fig. 6i). Both films have grown large OR2/t grains in which holes have hexagonal shape with circular edge-bumps, typical of a hole formed in a <111> single-crystal grain. The holes are scattered within the films with equivalent diameters of 5.32 ± 1.30 µm (Fig. 6b) and 4.94 ± 0.41 µm (Fig. 6i). Such similar hole sizes despite the difference in film thickness is likely to be fortuitous, related to the thermal acceleration of atomic diffusion in the thicker film annealed at a higher temperature.

The 270 nm thick film annealed at 1273 K (Fig. 6e) which has 10 µm wide OR1/t grains, has dewetted like a polycrystalline thin film [70]. Holes have irregular elongated non-convex shapes with discontinuous edge-bumps. This is related to the impingement of holes and/or to the extended grooving from the triple junctions along the grain boundaries of the film [52]. Holes have opened to sizes of 2.96 ± 0.86 µm (Fig 6e), smaller than the holes in the 200 nm / 1273 K film and in the 500 nm / 1423 K film. In both thinner films, the density of the holes is on the order of $10^4$ per mm² and their projected area covers 37 ± 3 % of the total surface of the films. This indicates that dewetting of a film by hole expansion along the grain boundary network is a slower process than the opening holes in a single-crystal grain.

Finally, the holes in the 670 nm thickest film annealed at 1423 K are more isotropic and surrounded by irregular ridges located further away from their edges. They are 2.5 times smaller than in the 500 nm thick film. While the total projected area of holes is similar (4 ± 1 %) in both the thicker films, the hole density is 10 times lower in the 500 nm thick film as reported in Table 3. This suggests that the grains in the 500 nm thick film have grown larger than the ones in the 670 nm thick one.

*3.3.3. Film break-up into isolated islands*

At 1423 K, the two thinner films have broken up into micrometer-sized facetted grains (see the inset of Fig. 6c) or elongated islands (Fig. 6c and 6f). This is an expected step following the percolation of holes like the one observed in the 270 nm film at 1273 K. In the 270 nm film, the finger-like islands are made of OR1 and OR1t grains with $\Sigma 3$ grain boundaries. The islands in the 200 nm film are mostly single-crystals, of the 4 possible ORs with a more isotropic shape than in the 270 nm film due to less material to diffuse to achieve equilibrated single-crystal grains [44]. Two break-up scenarios are observed depending on the extent of grain growth. (i) In the case of the 270 nm thick film, the grains are OR1 and OR1t and their sizes stagnate at 10 µm (see Table 2). Between these grains, the grain boundary groove faster than they move such that break-up propagates by grain boundary grooving from its textured state; (ii) In the case of the 200 nm thick film, the break-up of the film proceeds in two steps: like in the 270 nm films grooving first separates grains of the 4 different ORs, small OR1/t grains and larger OR2/t ones. The OR2 grains had time to grow before they get isolated by grooving, break-up into smaller OR2 grains by dewetting from their edges [71].

**4. Discussion**

**4.1. Film thickness and phase stability**

The fcc CoCrFeNi phase was found independently from film thickness, from 200 nm to 670 nm. The different grain growth and hole nucleation behaviors observed in the present work are clearly not related to any phase change since our CoCrFeNi films are fcc in the whole temperature range investigated: at room temperature (as deposited state) and when annealed from 0.55 $T_m$ up to 0.8 $T_m$. It is surprising that the as-deposited phase is fcc because according to the proposed phase diagram [6,17-21], the Cr-rich σ-phase is stable below 600 K. Actually, when the CoCrFeNi films are prepared on amorphous $SiO_2$/Si substrates, the σ-phase is observed in XRD in the as-deposited state (not shown here). Therefore, we suspect that the 3-fold crystallographic symmetry of the c-plane of sapphire, which favors the growth of <111>-fcc grains of preferred ORs and/or the high quenching rates during the process of sputter-deposition may kinetically suppress σ-phase formation.

**4.2. Preferred orientation relationship of the CrCoFeNi grains on c-sapphire**

The anisotropy of the film grain/substrate interface energy depends on the orientation of the grain relative to the substrate, i.e. the OR. Moreover, topographical features on the substrate surface (like steps) can participate in the growth of specific in-plane orientations [72].

The studies of the ORs of fcc pure metallic films on c-sapphire generally report that OR1/t is preferred [55,59,73,74] but OR2/t has also been observed [58,60,66,75]. Processing conditions (growth temperature, deposition rate, pressure of the sputter system) may play a role, as shown in [60]. It has also been reported that in samples prepared at high temperature, OR2/t is favored for Al [58,66], and it has been suggested that this preferred OR might be controlled by kinetics [58]. On the other hand, Curiotto et al. [55] have shown that when OR1/t Cu crystals formed by dewetting in the solid state are melted and then re-solidified, OR1/t is no longer preferred and the <110> in-plane direction in the Cu interface aligns on the <11$\bar{2}$0> in-plane direction of sapphire (i.e. the preferred direction of sapphire for alignment of the <110> direction of the OR2/t grains).

In this study, we found the same trend as in previous studies on the orientation relationships of fcc grains on c-sapphire: grains prefer to grow in OR1/t at low temperature but are likely to adopt

OR2/t at high temperature. However, OR2/t grains may not appear if they cannot expand from pre-existing small OR2/t grains [49]. In this work, CoCrFeNi films heated up to 0.8 $T_m$ either grow to 10 μm large grains in OR1/t or ~1000 μm large OR2/t grains within a ±10° in-plane deviation. Since the OR2/t grains have been found much less frequently in the as-deposited film (4 vol.%), it could happen that in the sample investigated, they are not observed after a high temperature anneal because they were not present in the as-deposited films (or in the measured region of the annealed film).

The interfacial energy may be lower for the OR2/t grains at temperatures above 973 K. DFT calculations performed at 0 K for the fcc pure Cu have shown that the interfacial energies of the OR1 and OR2 grains on sapphire are very close [76] but there is no information on how they evolve with temperature. Thus, it is not yet completely understood what leads to OR1/t or OR2/t. We suspect that the sapphire substrate steps may play a role on the stabilization of these two ORs, consistently with what has been shown in [72].

It is worth mentioning two recent papers in the literature that attempt to explore the impact of an alloying element on the microstructure of a metallic thin film in contact with an oxide plane. They investigated films of fcc binary alloys which were prepared by stacking of pure layers on single-crystal oxide planes (Ni-Cr on YSZ [77] and Ni-Fe on c-sapphire [78]). Although interfacial segregation of Cr [77] has been observed, there was no change of the preferred OR of Ni on YSZ since it was already dictated by the Ni layer that had been deposited in contact with the oxide crystal before Cr segregation occurred. The same situation stands for Ni-Fe alloy where Ni was also the as-deposited layer in contact with c-sapphire. In our experiments, the role of chemistry can be considered unbiased since the deposition method used ensured the chemical homogeneity of the 4-component films from the as-deposited state (confirmed by TEM-EDS studies of cross-sectional samples - not shown here). Interfacial segregation/adsorption and OR develop in parallel upon annealing.

One of the main conclusions of this paper is that despite the complex composition of the CoCrFeNi films, their annealed microstructure consists of grains of the same two preferred ORs as those of pure fcc metal films on the same substrate, in the same equivalent temperature range. This means that neither the lattice parameter nor the interfacial segregation of one or more of the alloy components is responsible for the preferred OR of a fcc phase on c-sapphire. The fact that the preferred ORs are identical for any fcc phase deposited on top of c-sapphire must be based on crystallographic properties of the substrate plane such as its interfacial steps, as shown in [72].

**4.3. Grain growth and annealed microstructure/morphology**

Here, the results are discussed by sorting the films according to the final OR of their grains (including their variants), i.e. the OR2/t-films and the OR1/t-films.

OR1/t grains are systematically found to have grown at the lowest annealing temperature of 973 K, while both OR1/t and/or OR2/t grains can be observed after anneals at the higher temperatures. When they have grown, the OR2/t grains are much larger than the OR1/t grains which means that the grain boundaries which delineate the OR2/t grains migrate faster.

In the OR2/t-films, grain growth has eliminated most of the grain boundaries upon annealing (above 973 K). Some triple junction grooves have had time, before being swept away by grain growth, to deepen down to the substrate, forming holes through the film. These holes are left isolated in one of the single-crystal OR2 or OR2t grains after the grain boundaries have detached from them under the driving force of grain growth. These isolated holes have a hexagonal shape typical of that of dewetting-holes within a {111} fcc single-crystal (see Figs. 6b and 6h). They are wider (for the same annealing duration) at higher temperatures, and in thinner films. When holes expand by surface diffusion, the film thickness enters only in a logarithmic term, such that its impact is limited to the short-time regime [52]. The other factors which determines the growth rate of holes are identical for all films; they are the equivalent contact angle of the film on the substrate, and the absolute value of the surface energy, as predicted by the 2D model of Zucker et al. [79] and as demonstrated by the experiments performed on Ni {100} single-crystals on MgO [71], on Si single-crystals on amorphous silica [68] or on Al large grains on c-sapphire [58,80]. Thus, the larger hole size observed for thinner films and higher annealing temperatures are likely due to the shorter time required for their formation by surface diffusion, a mechanism which is temperature activated and according to which the linear expansion of a groove is proportional to $t^{1/4}$ [51].

In the OR1/t-films, grain growth is much slower than in the OR2/t-films, and grains stagnate at sizes of about 10 μm, even at 1473 K. Thus, in the 270 nm thick films, it is likely that the apexes of the triple junction grooves reach the substrate before the grain boundaries have detached from them. This allows dewetting to propagate along the grain boundary network of the film [52,81]; at 1273 K, finger-shaped holes have formed as seen in Fig. 6e, and at 1423 K the holes have percolated into a network such that the continuous holey-film has switched to elongated islands as shown in Fig. 6f. In the thicker films, triple junction groove tips need more time to reach the substrate (proportional to $t^{1/4}$ [51]), and at 1423 K, grain growth, while slow, tends to eliminate the grain boundaries around the triple junction grooves before they become too deep and prevent the boundaries from moving. The triple junction grooves which have reached the substrate become isolated holes within one of the OR1 or OR1t grains. In the 670 nm thick films, the surface morphology that is observed is related to a transition from triple junction grooving with a shallow groove angle and distant bumps to dewetting-hole growth in a single-crystal with vertical walls and edge-bumps.

In the case of the 200 nm film annealed at the highest temperature, the film broke completely within a short time by grooving at the triple junctions and probably also at the grain boundaries. The resulting islands are crystals of which 25 % are OR2/t grains and the remaining OR1/t grains. This clearly shows that the OR2 grains were growing at the expense of the OR1/t grains before the complete rupture of the film.

Several studies have been carried out on grain growth upon annealing of pure metal films deposited on amorphous silica or on single-crystal silicon. They have shown that grain growth stagnates when grain size reaches 20 times the film thickness [53,64,65,82]. Larger grains exceeding 50 times the film thickness, have been found in ultrathin (< 100 nm) films of Si on $SiO_2$ [83]. Recently, much bigger grains have been prepared in fcc films (Ni and Cu) thicker than 300 nm, deposited on a ($1\bar{1}02$) sapphire substrate [49]. On this substrate, the Ni grain size could reach 300 times the film thickness when annealed at 0.82 $T_m$ (1423 K) for 1 h, and the Cu grain size could extend to 500 times the film thickness when annealed at 0.92 $T_m$ (1253 K) for 78 h. In the present work, CoCrFeNi grains well beyond these sizes have been found when their orientation relationship to the c-sapphire substrate is OR2/t and the film thickness and annealing temperature combination is such that grain growth is not impeded by grooving; after 1 hour at 0.8 $T_m$ (1423 K), the CoCrFeNi grains are more than 1000 times the film thickness. Stagnation of grain growth is attributed to the formation of grooves which prevent grain boundary motion [65]. We speculate that grooving takes place mostly on slow-moving grain boundaries. Thus, to overcome grain growth stagnation, the grain boundary migration rate must be high. Since we observe grain sizes exceeding 1000 times the film thickness, the grain boundary migration rate must be very large, i.e. their driving force and/or mobility must be high.

This observation disagrees with the statements found in literature that grain boundaries in MPEA alloys have a sluggish diffusion, because they are pinned by defects like dislocations [84] or because of a particular distribution of the components at the grain boundaries [85]. These arguments may apply for certain type of grain boundaries, but not for the <111> tilt boundaries of these films.

The adsorption of some of the component(s) of the alloy at grain boundaries has enhanced their mobility compared to the one for pure fcc metal like Al [58] at the same equivalent temperature. It has recently been shown by calculation that Cr adsorbs/segregates at one of the grain boundaries of the Cantor alloy (CoCrFeMnNi) [86]. Since this paper shows that only Mn could compete with Cr, it is likely that Cr also adsorbs at the grain boundaries of the quaternary CoCrFeNi. It is important to remind that adsorption/segregation at a grain boundary may either slow down (solute drag effect [87]) or accelerate (see for example [88]) grain growth. Thus, it looks like adsorption enhances mobility of certain of the grain boundaries of the <111>-textured CoCrFeNi alloy thin film on c-sapphire investigated in this paper.

Arguments based on energy decrease must be considered but they cannot be simply used to understand kinetics since reaching a lower energy state depends on the mechanisms involved in grain boundary motion. The two main energetics arguments are listed below:

(i) The interfacial energy anisotropy between sapphire and OR2/t grains vs. OR1/t grains could be enhanced by adsorption of some of the component(s) of the alloy at the interface, such that the interfacial energy with sapphire of the OR2/t grains has become much smaller than the one with the OR1/t grains at temperature higher than 973 K.

(ii) The grain boundaries between OR1 and OR1t are $\Sigma$3 boundaries while the grain boundaries between OR2 and OR2t deviate from $\Sigma$3 with an in-plane orientation spreading of about ±10° (also seen for Cu [66] and Al [58] on c-sapphire). The grain boundaries between OR1/t and OR2/t are general grain boundaries. Thus, from a strict crystallographic point of view, the energy of the OR2-OR2t grain boundaries should be larger than the one of the OR1-OR1t grain boundaries [89], and the energy of the OR1/t-OR2/t grains should be the largest. However, adsorption/segregation at grain boundaries in alloys is stronger at higher energy grain boundaries. It is probably larger at the OR2-OR2t and OR2/t-OR1/t grain boundaries than at the $\Sigma$3 grain boundaries between OR1 and OR1t. This may change the order of the grain boundary energies which decrease as adsorption increases.

In fcc phases, the {111} planes are the most densely packed and the lowest surface energy planes. Thus a <111> fiber texture minimizes the film surface and interface energies [83,89,90]. This is why this texture is usually found in thin fcc films [58,59,91]. During deposition of metallic films, surface energy driven grain growth can occur, and the resulting texture develop at temperatures as low as 0.2 $T_m$ [82]. Wong et al. [89] showed that during electron beam deposition of Au films at room temperature (about 0.22 $T_m$), the film develops a texture as soon as it becomes continuous. Chou et al. [90] have correlated deposition rates and textures of as-deposited thin Sb films. They have shown that when the deposition rate was < 2.5 nm/s, grains of the lowest surface energy were over-represented because atoms had time to diffuse. In the present work, after deposition at room temperature (0.18 $T_m$) at a relatively low rate of 0.13 nm/s, the films were already fcc and textured, with {111} planes parallel to the (0001) plane of the c-sapphire substrate surface, with grains of a preferred OR (OR1/t as shown on Fig 5). This metastable microstructure may have been stabilized by the 3-fold symmetry of the c-sapphire substrate.

This work shows that a complete understanding of grain growth in the CoCrFeNi <111>-textured film and more generally of bulk alloy, requires to investigate the reason why certain grain boundaries are accelerated while others are slowed down. This is a large piece of work which requests the fine characterization at the atomic scale of the fast and slow grain boundaries of the films and the track of their distribution change from the microstructure of as deposited <111>-textured films with nanometer grains, to the one of films with millimeter OR2/t grains.

## 5. Summary

This paper reports for the first time on the stability (phase, microstructure, morphology, grain growth and dewetting) of thin CoCrFeNi polycrystalline films deposited onto a single crystal substrate.

CoCrFeNi films were prepared at room temperature on c-sapphire by magnetron sputtering which allows control of a uniform composition. As-deposited films exhibit a nanocrystalline columnar fcc <111> fiber textured microstructure on the (0001) plane of sapphire. The columnar grains have sizes parallel to the surface plane of 30 nm to 100 nm for the deposited film thicknesses of 200 to 670 nm. The low deposition rate (0.13 nm/s) is one of the parameters responsible for this early texture formation, and the crystallographic symmetry of the c-plane of sapphire may be the reason for stabilization of a fcc structure of the deposited alloy.

The CoCrFeNi fcc phase was found to be stable from room temperature up to 1423 K. No other metallic phase precipitates. When Cr-rich regions appear, they are related to the formation of thin surface Cr oxide layers because of slight oxygen content in the annealing atmosphere.

The microstructure evolution of the film was investigated by tracking the density, size and shape of the grooves formed at the triple junctions between the grain boundaries upon annealing. Holes through the film nucleate at the grain boundary triple junction groove roots. Break-up is delayed in thicker films because more time is required for the triple junction groove to reach the substrate. Grain growth and grain boundary/triple junction grooving compete to control the morphology of the film upon annealing. In the OR2/t-films, grains grow to giant sizes, exceeding 1000 times the film thickness at 1273 K and above. In these films, holes which originated from former triple junctions become isolated within single-crystalline grains. Local dewetting takes place from these holes which expand with a hexagonal shape typical in <111> grains. In the films where OR1/t grains developed, the grain size is limited to few micrometers; these films are likely to dewet/break-up by propagation of the holes nucleated at triple junctions along their grain boundary network. Thus, the thinnest OR1/t films are unstable and break-up easily. The reason for the development of OR1/t or OR2/t is clearly connected to the annealing temperature and the presence of OR2/t seed-grains in the as-deposited film. Higher mobility of some grain boundaries leads to abnormal growth of OR2/t grains.

Finally, despite their complex composition, the films investigated behave similarly to pure fcc metal films on the same substrate. Neither the lattice parameter nor the segregation of one or several of the components of the alloy at the fcc/c-sapphire interface changes the final microstructure and texture. The fcc structure of the film is the only critical parameter controlling the annealed microstructure, while the complex chemistry plays on kinetics. Thus, quite unexpectedly, most of the conclusions on fcc film stability on c-sapphire found for these MPEA films can be extended to, or could be inferred from, pure metals.

**Acknowledgements**

This work was performed under the umbrella of the ANR-DFG joined project AHEAD. The French authors at CINaM and CEMEF wish to thank the Agence Nationale de la Recherche for support of their research under grant ANR-16-CE92-0015. The German authors at MPIE and RUB wish to thank the DFG for financial support within the projects DE 796/11-1 and LU1175/22-1.

All the authors wish to acknowledge Paul Wynblatt (CMU) and Céline Varvenne (CINaM) for useful discussions about grain boundaries in CoCrFeNi. Sapphire substrates were cut and cleaned in the PLANETE CT PACA cleanroom facility (CINaM). Benjamin Breitbach (MPIE) is acknowledged for XRD measurements and Stefan Hieke (MPIE) for discussions on dewetting, grain growth and metal-sapphire orientation relationships.